\documentclass{elsart41}
\usepackage{graphics}
\usepackage{graphicx}
\usepackage{amssymb}
\begin{document}

\begin{frontmatter}



\title{Magnetic Phases of Rare Earth Hexagonal Manganites}
%---- Don't remove this comment line! ----

\author[AA]{S. H. Curnoe\corauthref{Curnoe}},
\ead{curnoe@physics.mun.ca}
\author[AA]{I. Munawar}
%\author[BB]{C. Name3}

\address[AA]{Department of Physics and Physical Oceanography,
Memorial University of Newfoundland, St. John's NL, A1B 3X7, Canada}  
%\address[BB]{Institute of Chemistry, University of City2, Address2, Country2}

\corauth[Curnoe]{Corresponding author. Tel: 1-709-737-8888
fax: 1-709-737-8739}

\begin{abstract}

We describe the magnetic phases of hexagonal rare earth manganites
RMnO$_{3}$ using Landau theory.
A minimal model based on four one-dimensional magnetic order parameters is 
developed.

\end{abstract}

\end{frontmatter}

% main text

Hexagonal rare earth and yttrium manganites RMnO$_3$ (R = Ho, Er, Tm, Yb, Lu, Y or Sc) belong to an unusual class of materials known as ``multi-ferroics", which
display simultaneously electric and magnetic ordering.  
Most of the hexagonal manganites are ferroelectric below
a very high temperature (T$_{FE}\approx$ 900 K) and order magnetically
at a lower temperature (T$_{N}\approx$ 100 K).
A complex
phase diagram involving several
different magnetic order parameters has been investigated using
second harmonic generation \cite{shg1,fiebig1} and magnetic and heat capacity
measurements
\cite{lorenz1}.   The partially filled 4f shell of the
Ho, Er, Tm and Yb compounds has been implicated in low temperature 
region of the phase diagram
\cite{lonkai1}.

Symmetry plays an important role in the analysis of these systems.
Group theoretical methods may be used to:
{\em i)} classify spin structures that contribute to 
magnetic ordering; {\em ii)} determine coupling between magnetic order 
parameters and applied fields, including magneto-electric coupling;
and {\em iii)}  find phase diagrams derived
from Landau free energy expansions.
In this article we discuss preliminary results based on analysis of the Landau
free energy.

Below T$_{FE}$, the space group of hexagonal RMnO$_3$ is
\#185, P6$_3$cm.  The corresponding point group  C$_{6v}$ has 
four one-dimensional representations A$_{1,2}$, B$_{1,2}$, and 
two two-dimensional representations E$_{1,2}$.
These labels are also be used to denote the representations of the 
magnetic point group which correspond to magnetic order parameters.
Multiple copies of every order parameter are possible for the spins of
both Mn and rare earth ions.  
%We will not consider a microscopic description of the order parameters
%in  this article, instead we will concentrate only on the
%symmetry properties.   
There is an antiferromagnetic spin configuration 
involving Mn spins in the $xy$-plane for each 1D order parameter (see
Ref.\ \cite{fiebig1}).
Additional spin configurations aligned with $z$ for
both Mn and R spins yield other 1D order parameters, as listed below.
In these cases, 
the A$_2$ order parameter is associated with ferromagnetic ordering.
Spin configurations for the 2D
order parameters $E_{1,2}$ are also allowed, but so far there is
no evidence that they appear in the phase diagram.
%These involve Mn spins in the $z$-direction and the $xy$ plane and
%Ho spins in the $xy$ plane.
\begin{table}[h]
\caption{1D order parameters for spin configurations aligned with $z$.}
\begin{tabular}{ccl}
\hline
R (2a) & 
A$_2 \; \;$ & $S_{1z}+S_{2z}$\\
& B$_1$ & $S_{1z}-S_{2z}$\\
\hline
R (4b) & 
A$_1$ & $S_{1z}+S_{2z}-S_{3z}-S_{4z}$\\
& A$_2$ & $S_{1z}+S_{2z}+S_{3z}+S_{4z}$\\
& B$_1$ & $S_{1z}-S_{2z}+S_{3z}-S_{4z}$\\
& B$_2$ & $S_{1z}-S_{2z}-S_{3z}+S_{4z}$\\
\hline
Mn (6c)
& A$_2$ & $S_{1z}+S_{2z}+S_{3z}+S_{4z}+S_{5z}+S_{6z}  $\\
& B$_1$ & $S_{1z}+S_{2z}+S_{3z}-S_{4z}-S_{5z}-S_{6z}  $\\
%& A$_1$ & $-(S_{1x}-S_{4x}+\frac{1}{2}(S_{2x}-S_{5x}+S_{3x}-S_{4x})
%+\frac{\sqrt{3}}{2}(-S_{2y}
\hline
\end{tabular}
\end{table}
%In this article we present a minimal Landau theory which describes the
%magnetic phase diagrams of (Ho, Er, Tm, Yb)MnO$_3$.
The presence of 1D order parameters yields the corresponding magnetic space 
groups
P6$_3$cm (A$_1$), P6$_3$\b{c}\b{m} (A$_2$), 
P\b{6}$_3$c\b{m} (B$_1$), and P\b{6}$_3$\b{c}m (B$_2$).

We begin by discussing the generic features of the magnetic phase diagrams
of [Ho,Er,Tm,Yb]MnO$_3$, as shown in Refs.\ \cite{fiebig1,lorenz1}.
%Only the one-dimensional order parameters listed above are present.
In all 
compounds, for $H_z=0$, the first magnetic phase to appear as a function of
decreasing temperature is B$_2$.  The
A$_2$ phase appears at lower temperatures and non-zero applied
fields, separated from the B$_2$ phase by a broad region of hysteresis.
The A$_2$ phase appears at zero field only in ErMnO$_3$.  
Additional phases A$_1$ and B$_1$ appear at low temperatures and
fields in HoMnO$_3$ only.  In this case, an intermediate phase 
between B$_2$ and B$_1$ has been found \cite{lorenz1}.

The order parameters for the phases A$_{1,2}$ and   B$_{1,2}$ 
are denoted by $\eta_{1,2,3,4}$ respectively.
The minimal model which describes the A$_2$-B$_2$ phases is
$$
F = \alpha_2 \eta_2^2 + \beta_2 \eta_2^4 + \alpha_4 \eta_4^2 
+\beta_4 \eta_4^4 + \gamma_{24}\eta_2^2\eta_4^2
$$
$$ 
- H_z(\rho_1 \eta_2  +\rho_2 \eta_2^3 + \rho_3 \eta_2 \eta_4^2)
$$
where $\alpha_i$, $\beta_i$ and $\gamma_{ij}$ and $\rho_i$
are phenomenological coupling constants and $H_z$ is the magnetic
field along the $c$-axis.
$\alpha_i$ are temperature dependent and
$\beta_i > 0$
% and  $\beta_i + \beta_j + \gamma_{ij} >0$
is required for stability. 
In all four compounds $\alpha_4$ changes sign at $T_N \approx 80$K.
Note that $\eta_2$ transforms in the same way as $M_z$. 
%Terms involving higher orders in $H_z$ are also permitted by symmetry.

This simple model can account for all of the features so far observed.
In zero applied field, the model allows for four different phases:
$(0,0)$ (the parent phase),
which may be connected to either $(\eta_2,0)$ (A$_2$-phase), $(0,\eta_4)$ (B$_2$-phase)
or $(\eta_2,\eta_4)$ (a mixed phase) by second order phase transitions.
These are found by solving the set of coupled equations
$\partial F/\partial \eta_i = 0$, subject to the 
minimisation
conditions $(\partial^2 F/\partial \eta_2^2)>0$ and $(\partial^2 F/\partial \eta_2^2)(\partial^2F/\partial\eta_4^2)
-(\partial^2F/\partial\eta_2\partial\eta_4)^2>0$.
The mixed phase can be a 
minimum of $F$ only when $4\beta_2\beta_4>\gamma_{24}^2$.  Its
existence is {\em not}
the result of hysteresis.   
The model also allows for the coexistence of two or more
different phases by hysteresis.
Experiments suggest that in fact there is a 
first order transition between the B$_2$-phase and the A$_2$-phase in all
materials with a broad area of hysteresis between, therefore we conclude
that $\gamma_{24}^2 > 4\beta_2 \beta_4$.  Then stability requires that
$\gamma_{24}>0$.  Anomalies in the $c$-axis magnetisation at the B$_2$ phase
boundary \cite{lorenz1}  are evidence that $B_2$ and $A_2$ are coupled.
%In finite applied field the model predicts that
%$\eta_2$ grows linearly with $H_z$.  
%This is not observed experimentally,
%which suggests that the coupling $\rho_1$ is small.  
In general  $\eta_2$   grows linearly with applied field
but it will still be subject to a transition 
in the sense that a change in sign of $\alpha_2$ will increase the number
of minima of the Landau functional.
%The field dependence of the phase transition line to A$_2$ 
%implicates non-linear
%couplings to the applied field.

The low temperature part of the phase diagram of HoMnO$_3$ contains
several additional phases.
Neutron diffraction studies \cite{lonkai1}
indicate that the new phases $B_1$ and $A_1$ which appear below 35K are
due ordering of the Ho spins, presumably associated with the rather large
magnetic moment of the Ho ion.
Additional order parameters $\eta_1$ and $\eta_3$ are required to 
describe the phases in HoMnO$_3$.   New terms in the free energy 
take precisely the same form as those involving $\eta_2$ and $\eta_4$,
as well as a term of the form $\eta_1\eta_2\eta_3\eta_4$.
Field dependent terms are found by replacing $\eta_4$ with $\eta_1$ and
$\eta_3$ in the field dependent terms of $F$, and a term of the form $H_z\eta_1
\eta_3\eta_4$ is also allowed.
This model may be solved exactly in zero applied field, and it 
is found that the allowed phases are $(0,0,0,0)$ (the parent phase),
$(\eta_1,0,0,0)$ {\em etc.} (A$_i$ or B$_i$) and $(\eta_1,\eta_2,0,0)$ {\em etc.}
(mixed phases involving two order parameters). 
In addition, hysteresis effects will increase the complexity of the 
observed phase diagram.
%(and in general $\eta_2$ will be non-zero in an applied field.

Experimentally, 
%a broad hysteris between the B$_2$ and A$_2$ phases is observed \cite{fiebig1},
a new intermediate phase between B$_2$ and B$_1$ is found at low
temperatures \cite{lorenz1}.  This phase may be a true minimum of $F$
of the form $(0,0,\eta_3,\eta_4)$
 with symmetry P\b{6}$_3$.  However, the most recent results 
show hysteresis {\em within} the intermediate phase region,  which
suggests that this phase may actually be further subdivided \cite{yen1}.  
%The actual magnetic symmetry of the intermediate phase has not yet been
%measured directly.
%but no true mixed phases of any combinations are found,
%so in general we expect that $\gamma_{ij} > 4\beta_i\beta_j$ 
%where $i$ and $j$ index phases which are neighbours on the phase diagram.

Finally, we comment on the magneto-electric effect in HoMnO$_3$.
Linear coupling of magnetic and 
electric field of the form $\alpha_{ij}E_iH_j$ can only occur in a subset
of magnetic point groups 
in which both inversion symmetry and time reversal symmetry are broken.
As noted in \cite{lottermoser1}, magneto-electric coefficients
$\alpha_{zz}$ and $\alpha_{xx}=\alpha_{yy}$ may be non-zero in P6$_3$\b{c}\b{m}
(A$_2$ order),
while $\alpha_{xy} = -\alpha_{yx}$ in P6$_3$cm (A$_1$ order).
Other coefficients may exist in the presence of E$_1$ or E$_2$ 
magnetic order parameters, if they occur.
Magneto-elastic coupling has been observed \cite{delacruz1}, but this
is secondary to the magneto-electric effect, 
since strain alone cannot induce linear
coupling between magnetic and electric orders.
In HoMnO$_3$, evidence of magneto-electric coupling from
dielectric anomalies was found in the intermediate B$_1$-B$_2$
region \cite{lorenz2}.  A mixture of B$_1$ and B$_2$ order parameters,
which would yield P\b{6}$_3$,  does
not lower the symmetry enough for magneto-electric coupling.  
Symmetry lowering in the vicinity of domain walls has been proposed
\cite{lottermoser2}.   Another possibility is that the magneto-electric effect
is due
to A$_1$ domains, since the A$_1$ phase region is adjoined to 
regions of hysteresis within the intermediate phase \cite{yen1}.

%To summarise, we have considered various aspects of magnetic
%ordering and symmetry in RMnO$_3$ compounds.

%
%
%\begin{figure}[!ht]
%\begin{center}
%\includegraphics[width=0.45\textwidth]{fig1}
%\end{center}
%\caption{This is the first picture}
%\label{fig1}
%\end{figure}
%
%

%\vspace{.04in}
%\section*{Acknowledgement}
We thank I. Sergienko for many helpful discussions.  
This work was supported by NSERC of Canada.

\end{document}